\documentclass[%
 aip,
 amsmath,amssymb,
 reprint,%
]{revtex4-1}
\usepackage{float}

\usepackage{graphicx}
\usepackage{svg}
\usepackage{dcolumn}
\usepackage{bm}

\usepackage[utf8]{inputenc}
\usepackage[T1]{fontenc}
\usepackage{mathptmx}
\usepackage{etoolbox}
\usepackage{array}
\usepackage{tabularx}

\makeatletter
\def\@email#1#2{%
 \endgroup
 \patchcmd{\titleblock@produce}
  {\frontmatter@RRAPformat}
  {\frontmatter@RRAPformat{\produce@RRAP{*#1\href{mailto:#2}{#2}}}\frontmatter@RRAPformat}
  {}{}
}%
\makeatother
\begin{document}

\preprint{AIP/123-QED}

\title[Ablation Removal of Transport-Blocking Defects in Surface-Electrode Ion Traps]{Ablation Removal of Transport-Blocking Defects in Surface-Electrode Ion Traps}
\author{T. Maddock}

\author{P. Rahimi}%
 \email{W.K.Hensinger@sussex.ac.uk}
 
\affiliation{ 
These authors contributed equally to this work
}%

\affiliation{ 
Sussex Centre for Quantum Technologies, University of Sussex, Brighton, BN1 9RH,
U.K.
}%
\author{M. Aylett}
\affiliation{ 
Sussex Centre for Quantum Technologies, University of Sussex, Brighton, BN1 9RH,
U.K.
}%

\author{R. Barcan}
\affiliation{ 
Sussex Centre for Quantum Technologies, University of Sussex, Brighton, BN1 9RH,
U.K.
}%

\author{S. Weidt}
\affiliation{ 
Sussex Centre for Quantum Technologies, University of Sussex, Brighton, BN1 9RH,
U.K.
}%

\affiliation{ 
Universal Quantum Ltd, Brighton, BN1 6SB, U.K.
}%
\author{W.K. Hensinger}
\affiliation{ 
Sussex Centre for Quantum Technologies, University of Sussex, Brighton, BN1 9RH,
U.K.
}%
\affiliation{ 
Universal Quantum Ltd, Brighton, BN1 6SB, U.K.
}%
\date{\today}

\begin{abstract}
We demonstrate \textit{in situ} removal of a transport-blocking defect on a surface-electrode ion trap device using a Q-switched Nd:YAG 532 nm pulsed ablation laser. This approach eliminates the need to vent and rebake the vacuum system, providing a low-overhead defect-remediation technique well suited for ion-shuttling architectures where system modifications typically incur substantial downtime - particularly in shuttling focussed experiments operating at temperatures that necessitate bakes. Additionally, the hardware used is readily available in many ion trap laboratories, making this solution attractive to experiments operating in such regimes. Following ablation, we observe near-unit shuttling success rates across the previously obstructed region and measure micromotion levels that remain within acceptable limits. This technique enables rapid, reliable restoration of transport pathways without interruption to experimental operation.
\end{abstract}

\maketitle


Reliable ion transport is an essential requirement for many proposed architectures for trapped ion quantum information processing. Surface-electrode traps increasingly incorporate
complex junctions, long shuttling pathways, and multi-zone architectures to move
ions between memory, interaction, and detection regions~\cite{winnipaper,
Kielpinski2002, lekitsch2017blueprint, ferdi_paper, wunderlich_paper, Seidelin2006, Chiaverini2005, Schindler2013}. In such architectures, the integrity of transport links directly
impacts computational fidelity, system reliability and information throughput: high-fidelity shuttling
enables fast algorithmic execution~\cite{Bowler2012}, multiplexed
operations~\cite{Blakestad2009}, and modular integration across separate trap
chips or cryogenic stages~\cite{matterlink, Nigmatullin2021}. Even a single
obstruction can halt all downstream operations, making uninterrupted transport
channels essential for large-scale designs.

Surface-trap defects can arise through multiple mechanisms; most notable are particulate
contamination due to neutral atom flux exposed to the chip during ion loading\cite{zhang2020convenient, Brownnutt2015, Hite2012}, debris from electrode edges, and redeposition
from laser-driven photoionization or ablation sources~\cite{Sedlacek2018,
Shu2014}. Debris 10s of micrometres in size can introduce strong local
perturbations to the pseudopotential, generating excess micromotion, induce static offsets via beam charge up and
completely block transport through narrow junctions. Typical remediation of such problems introduce long experimental downtime, owing to long bakeout times, which becomes
increasingly impractical as experiments scale into multi-module or cryogenic
architectures~\cite{Wang2021, Monroe2014}. Minimising vent–bake cycles is therefore a major
engineering priority.

These constraints have motivated interest in \emph{in situ} laser-cleaning techniques. Prior work has demonstrated ultraviolet pulsed-laser cleaning to remove contaminants and reduce anomalous heating~\cite{Allcock2011, Hite2012}, continuous-wave laser cleaning to modify adsorbates~\cite{Daniilidis2014}, and laser desorption methods for electrode surface recovery~\cite{Czaplewski2017}. While these studies show that targeted laser illumination can modify electrode surfaces without breaking vacuum, we note that these techniques have been used to remove adsorbed atomic and molecular layers from surfaces. By contrast, far less attention has been devoted to the removal of large, discrete, transport-blocking particulates, despite their severe operational consequences. This gap motivates the development of fast, reliable, \emph{in situ} laser-based methods capable of restoring blocked transport pathways without interrupting system operation.

\begin{figure}[H]
\includegraphics[width=0.5\textwidth]{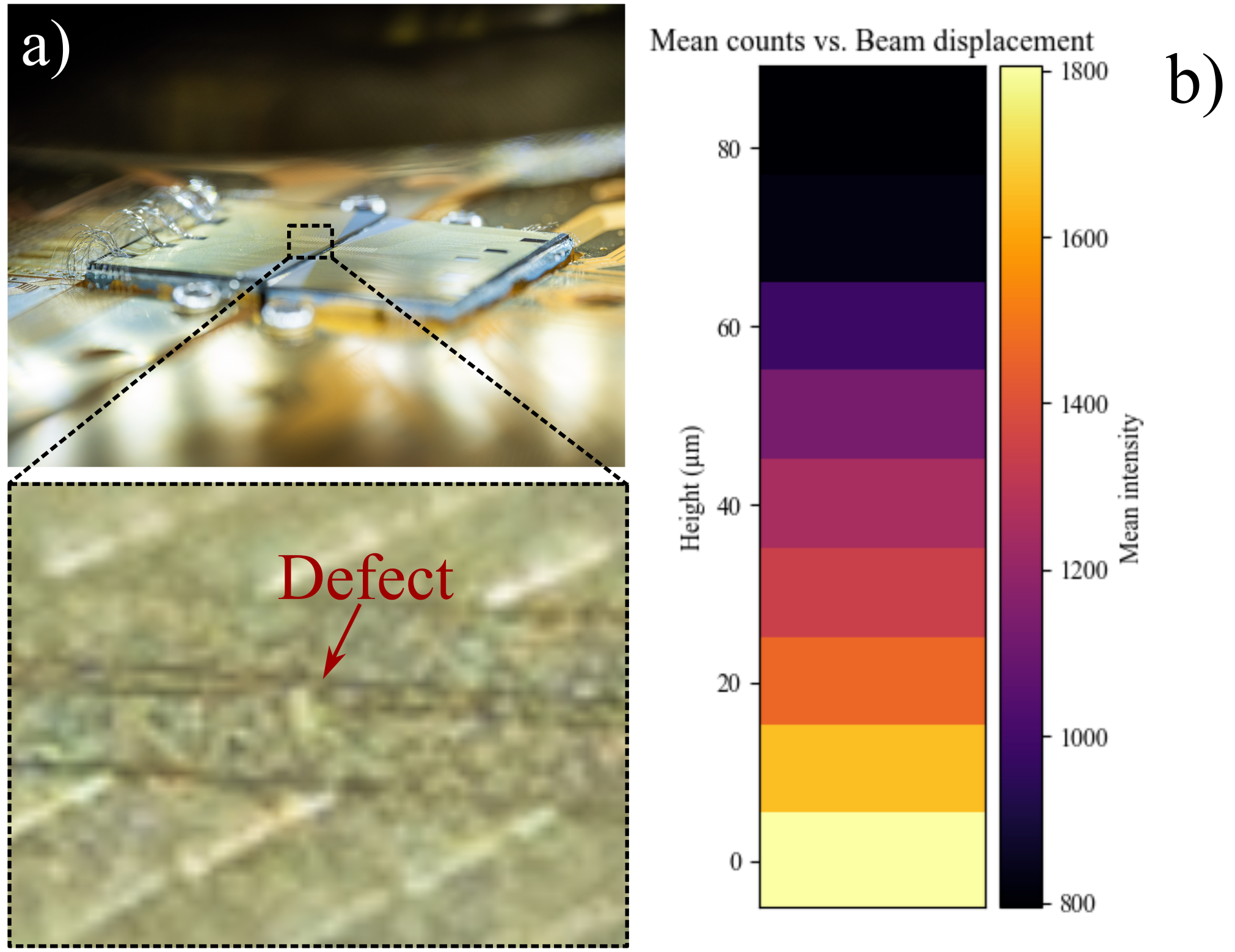}
\caption{ \label{fig:defect}a) Side-view image of the transport-blocking defect on the ion trap chip. The inset shows a magnified view of the defect region. b) EMCCD pixel data corresponding to a region of interest (ROI) centred on the defect as a guide laser illuminates it, and scanned perpendicular to the chip surface. The measurement is consistent with height estimates based on high resolution images, indicating a defect height of approximately 65~\textmu m.}

\end{figure}
In this study, a two-module quantum processor experiment was impeded by a transport blocking defect. The location proved to be particularly egregious due to its position between the gate zone of one of the chip modules and the second module, disallowing shuttling thermometry measurements. 

Highlighted in Fig.\ref{fig:defect}, analysis of high-resolution images and reconstruction using a scanned guide laser allowed us to estimate the dimensions of the defect at approximately $65\,\mu\mathrm{m} \times 40\,\mu\mathrm{m}$, confirming that it was large enough to block ion transport and detected during routing electron microscope detection, which implies that the origin of the contaminant was following the bake out procedure.

To address this, we implemented an \emph{in situ} removal technique using a Yb-doped 1.5~ns pulsed laser, carefully aligned and focused to the defect. This approach avoids venting the vacuum system or performing a full bakeout, eliminating the long downtime and alignment risks associated with conventional cleaning methods. The method enabled complete restoration of transport through the previously blocked region while minimizing the risk of collateral damage to nearby electrodes.
 
The ion trap assembly is comprised of two spatially disparate surface ion trap modules\cite{matterlink}, cryogenically cooled to typical operating temperatures of 47 - 50 K. 

\begin{figure}
\hspace{-5mm}

\includegraphics[width=0.52\textwidth]{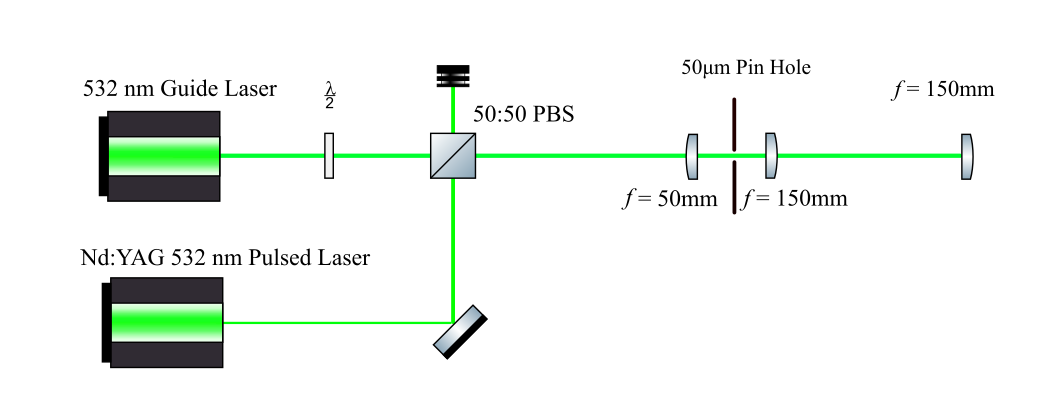}
\caption{\label{fig:optics} The beam path taken by both the guide laser and the ablation beam. The telescope serves to guarantee precise overlap of the ablation and guide beam, ensure sufficient beam quality to minimise risk to other areas of the chip, and to expand the beam prior to the final lens }
\end{figure}

Ablation is performed using a 532nm, 1.5ns pulsed laser, with a maximum available pulse energy of 2mJ. As indicated in Fig.\ref{fig:optics}, precise alignment to the microchip prior to ablation pulses necessitated the use of a 532nm continuous wave guide laser, with a comparatively  lower peak fluence.

\begin{figure}
\includegraphics[width=0.5\textwidth]{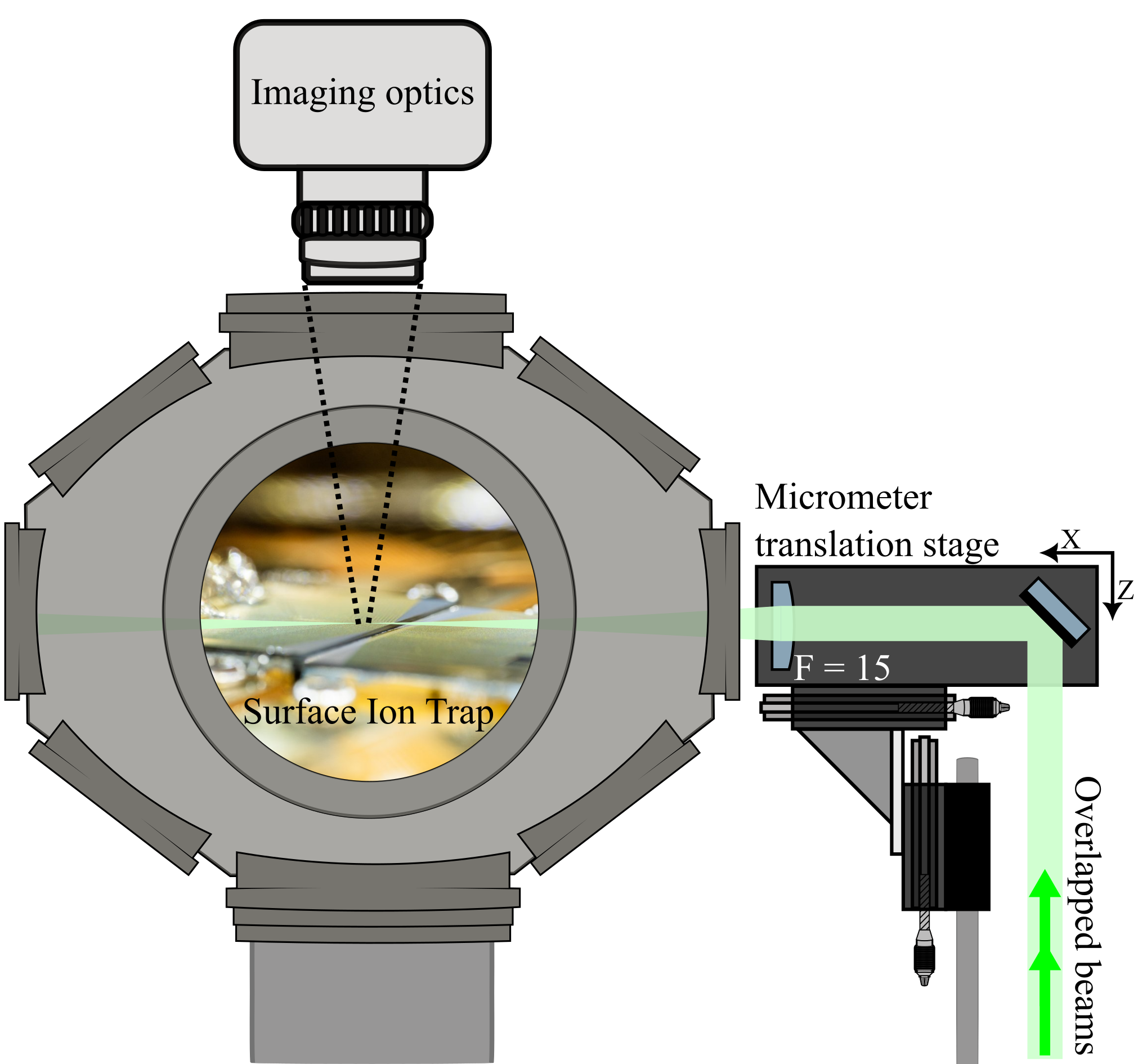}
\caption{Schematic diagram of the UHV system housing the surface ion trap, together with the imaging and laser ablation setup. The combined guide and ablation laser beams are delivered through a side window into the UHV chamber. Both beams operate at the same wavelength (532~nm) to enable reliable alignment and to avoid chromatic aberrations that could lead to systematic misalignment. Beam alignment is performed in the \(XZ\) plane using a micrometer translation stage, while focusing is achieved with a lens of focal length \(F = 150\).}
\end{figure}

\begin{figure*}
\includegraphics[width=1.0\textwidth]{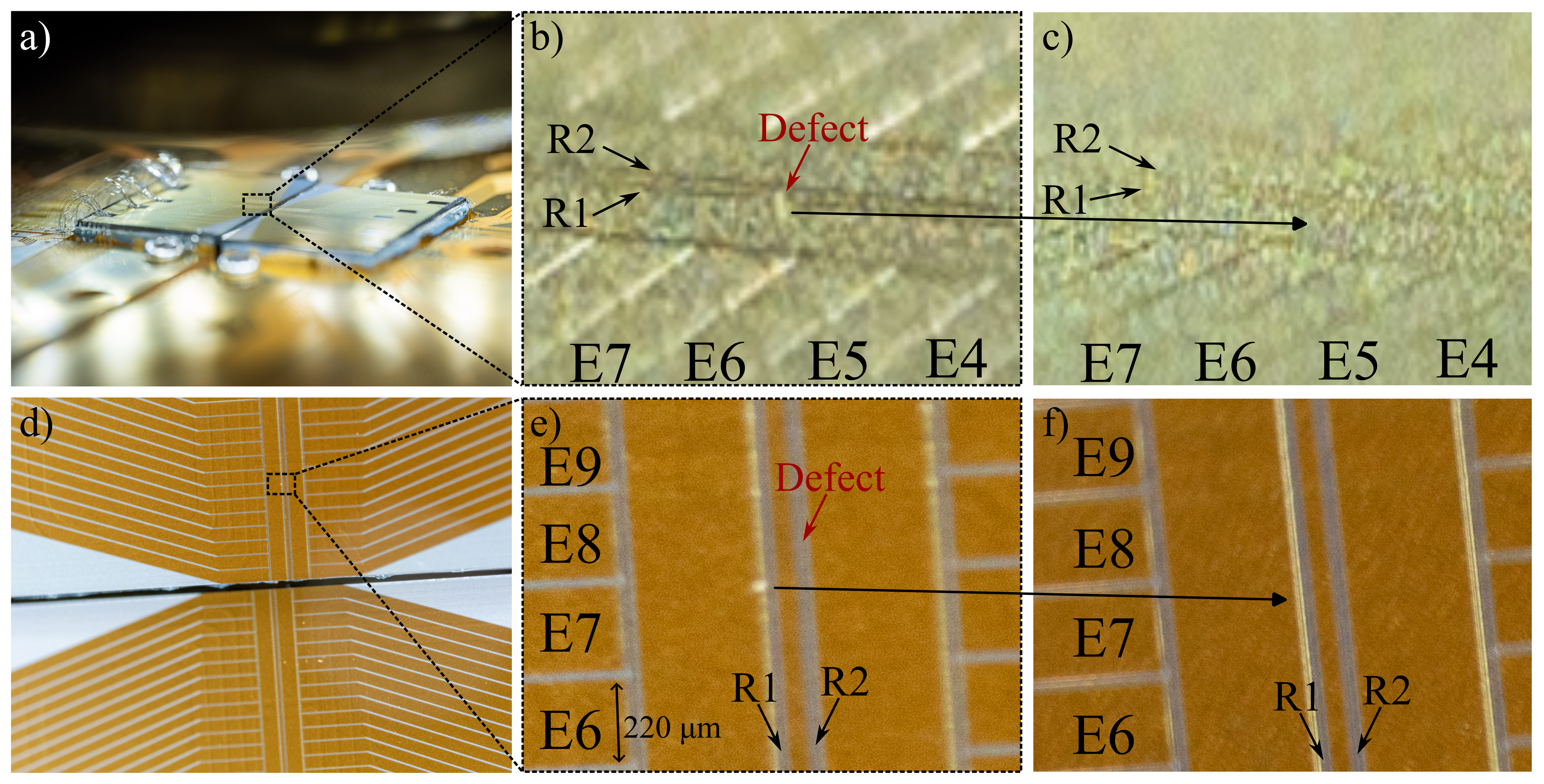}
\caption{Images of the ion trap chip acquired through the side windows a)-c) and the top windows d)-f) of the vacuum chamber. Panels a) and d) show an overview of the chip region containing the defect, while panels b), c), e), and f) provide magnified views of the defect location and the chip surface after defect removal. Electrodes are numbered sequentially starting from the central gap. The defect is located at the edge of electrode 7. The electrode width is indicated for scale.}
\end{figure*}

The ablation and guide lasers are combined using a polarising beam splitter, both of which are then passed through a beam expanding telescope with a 50um pin hole to ensure both beams are sufficiently Gaussian. Beam expansion ensured a small spot size following the final alignment to the defect. Both beams are carefully overlapped prior to precise chip alignment. The combined beams are then steered upwards to a motorised translation stage wherein beams are focussed to the defect using an adjustable 150mm lens. The motorised stage enables $\mu$m steps in vertical translation, allowing for sufficient precision for pulse targetting, given the 65$\mu$m approximate height of the defect.
To reduce risk to neighboring electrodes, and electrode wirebonds, the beam was expanded to a large beam waist of approximately 0.856mm; the lens prior to the vacuum chamber was adjusted to focus the beam to exactly the defect's location, such that the beam's fluence rapidly declined at all points other than the target. With the guide laser used to illuminate solely the defect, the beam was then carefully parallelized with respect to the chip thus minimising electrode exposure to the beam. Additionally, pulses were applied in short bursts, with 200ms between pulses, such that continuous heating of the chip could be avoided, and solely ablation of the targeted defect could be achieved. The characteristic thermal diffusion time may be estimated as
$t_{\mathrm{diff}} \sim L^{2}/\alpha$, which for gold
($\alpha \approx 1.3\times10^{-4}\,\mathrm{m^{2}\,s^{-1}}$) yields diffusion
times of $\lesssim 1\,\mathrm{ms}$ even over $100\,\mu\mathrm{m}$ length
scales, ensuring that the 200~ms interpulse delay is more than sufficient
for complete thermal relaxation between successive ablation pulses. Given the empirically determined ablation thresholds of materials within the vacuum chamber (Table 1), we estimate the laser fluence incident on surrounding components. Here, the fluence is defined as the laser energy per unit area, while the ablation threshold corresponds to the minimum fluence required to induce material removal. To provide a conservative estimate, we model the beam as having a 40 $\mu m$ diameter at a distance of 60 $\mu m$ above the trap surface, allowing for imperfect alignment. In this context, “conservative” denotes that the beam height was aligned to exceed this, but we allow for imprecision in the alignment procedure. This approach enables assessment of whether surrounding materials are exposed to fluences approaching or exceeding their respective damage or ablation thresholds.

\begin{table}[htbp]
\small
\caption{\label{tab:ablation_bulk_refs}
Representative single-pulse laser ablation fluence thresholds for bulk metals used in surface-electrode ion traps.}
\begin{ruledtabular}
\begin{tabular}{ll}
Material & Threshold and reference range \\
Au & $\sim$1--4 J/cm$^2$; 5--10 ns, bulk Au \cite{Torrisi2004,NS_ablation_bulk_Au} \\
Al & $\sim$2--8 J/cm$^2$; 5--10 ns, bulk Al \cite{Vladoiu2008, NS_ablation_Al_bulk} \\
Steel & $\sim$0.1-0.3 J/cm$^2$; 5 ns, bulk stainless steel \cite{Wisse2012} \\
\end{tabular}
\end{ruledtabular}
\end{table}

It is apparent that the gold electrodes themselves are most at risk, given that with peak fluences as high as $7 J/cm^2$ exposed to the defect, with a beam diameter of 40$\mu m$, the fluence incident on the chip surface is approximately $1.9 \times 10^{-3} J/cm^2$.

By sweeping the guide laser up and down with respect to the defect and the trap individually, recording pixel data using an EMCCD camera, an approximate height of the defect could be ascertained; such that following a series of ablation pulses, an assessment of any change in the defect's shape could be judged.

Shuttling waveforms were generated using sequential least squares programming (SLSP), based on modelled FEM data of the two modules \cite{matterlink}. Axial transport voltages were applied to shuttle ions from electrode~7 to the centre of electrode~9 (see Fig.~\ref{fig:umotion}), corresponding to an electrode separation of 220$\mu$m and a transport step size of 3$\mu$m. Transverse displacement of the ion was achieved by applying static voltages asymmetrically to the rotation rails -- extraneous electrodes not involved in static well simulation, parallel to the direction of transport - with opposite polarity applied to rotation~A and rotation~B, resulting in a controlled shift of the ion position perpendicular to the transport direction within the plane of the trap surface. 

Because the 369nm cooling laser propagates parallel to the chip surface at an angle of 45$^\circ$ relative the rotation rails, in-plane micromotion could be characterised using standard photon-correlation techniques \cite{Berkeland1998, Keller2015}. Micromotion compensation was performed at each axial position along the shuttling trajectory, enabling a spatially resolved assessment of distortions in the RF pseudopotential induced by the transport-blocking defect. The resulting compensation patterns are shown in Fig.~\ref{fig:umotion}. 

\begingroup
\setlength{\parskip}{0pt}  
\setlength{\parindent}{1em} 
\begin{figure}[]
\hspace{-5mm}
\includegraphics[width=0.48\textwidth]{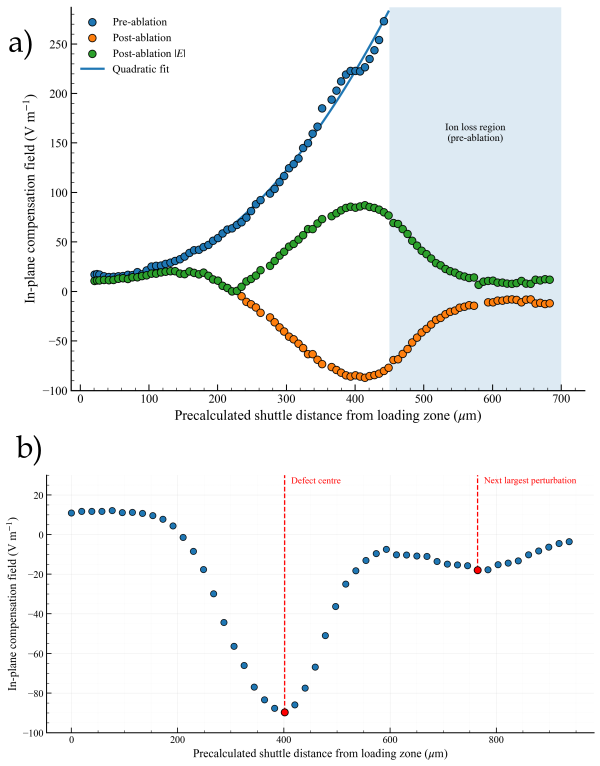}
\caption{\label{fig:umotion} a) Micromotion compensation voltages required to zero photon correlation modulation indices to zero. Yellow denotes the quadratrically expanding voltages required to compensate prior to defect removal - further data points could not be collected due to guaranteed ion loss, as denoted by the blue shaded region. Green denotes the micromotion voltages required following removal of the defect. b) Sampling further along the trap, highlighting the next maxima of compensation force required to zero photon correlation signals.}
\end{figure}
A threshold for complete removal of the defect was found by incrementally increasing the pulse power of ablation pulses from 10 to 80 percent, which corresponds to beam fluences at the target location from 0.56 $J/cm^2$ to $6.8 J/cm^2$. At a fluence of $5.6,\text{J/cm}^2$, complete removal of the defect was observed. This was initially indicated by the absence of scattering from the guide laser incident on the defect, and subsequently confirmed through high-resolution imaging. Assuming an ideal, gaussian beam, the fluence incident on neighbouring electrodes at the beam's focal point was $1.9 \times 10^{-3} J/cm^2$ at the highest fluence exposed to the defect during trials - much lower than empirically found ablation thresholds for gold seen in the literature. Assuming the focal point was exactly the defect's location, and that the beam was indeed perfectly parallel to the trap surface, this indicates that the fluence exposed to the trap could only have been lower than this estimated value, highlighting the minimal risk of catastrophic damage to the trap. Using a high magnification lens, images through both viewport A and B showed no obvious signs of the defect's residue, which agreed strongly with observations noted via the EMCCD camera. 

Following the ablation run, trapping was possible shortly afterwards. Shuttling past the location previously obstructed by the defect became immediately possible, with no empirical characterization of RF perturbation and dynamic compensation. Over 22500 trials, shuttling round trips were successful every time, implying that the round - trip shuttling error rate $ \leq1.3 \times 10^{-4}$, while before ablation shuttling was found to be unsuccessful in every trial out of $>300$. Additionally, by manually  increasing DC voltages active on electrodes neighbouring a trapped ions' location and monitoring ion perturbation, it could be seen that no DCs were disconnected or shorted following the ablation run.

To quantify residual perturbations of the RF pseudopotential following the ablation run, we measured the ion’s micromotion four days post-ablation (Fig.~\ref{fig:umotion}). The ion’s continued survival enabled a high-resolution assessment, with compensation data collected at each step along the shuttling trajectory. Analysis of the residual micromotion along the trap axis implies the presence of a correlated, localized deformation in the trapping potential. Prior to ablation, the required compensation force increases quadratically, eventually leading to the inevitable loss of the ion. Using the measured in-plane displacement vectors and the local electrode voltage-to-position Jacobian, we estimate that the maximum ion displacement induced by this RF distortion is approximately \(1.5~\mu\text{m}\), consistent with the range of displacements typically observed during standard shuttling operations. Sampling further along the trap, the next-largest compensation field required to null the micromotion in this plane was \(18.177(1)~\text{V m}^{-1}\), compared with a peak compensation field of \(88.95(9)~\text{V m}^{-1}\) in the vicinity of the defect. Though statistically significant, the perturbation at the location of the defect warrants compensation voltages of -0.3036V, which remains within typical bounds seen within this lab. This strongly suggests the presence of a shallow crater-like feature at the defect site, where material has been removed.

We have demonstrated a rapid and effective in situ method for removing a transport-blocking defect from a surface-electrode ion trap using nanosecond-pulsed laser ablation at 532 nm. The technique restores ion transport without venting the vacuum system or performing a bakeout, substantially reducing experimental downtime and risk. Following ablation, near-unity shuttling success was achieved across the previously obstructed region, while residual micromotion remained within acceptable limits and could be compensated using standard procedures.

The defect was removed without observable damage to surrounding electrodes or degradation of trapping performance, indicating that pulsed laser ablation can be applied selectively and safely to localized surface features. Because the approach is compatible with cryogenic operation and requires no modification of the trapping apparatus, it is well suited to shuttling-based and modular ion trap architectures where access for repair is limited.

This work establishes in situ laser-based remediation as a practical tool for maintaining transport reliability in complex trapped-ion systems, and highlights its potential to improve long-term stability and uptime as ion trap platforms continue to scale. More broadly, this technique could be extended to repair, \textit{in situ}, other quantum devices kept in vacuum that rely on microfabricated electrodes. For instance, superconducting quantum circuits and hybrid optomechanical systems may benefit from localised removal of defects or contamination that degrade device performance, particularly in cryogenic or ultra-high vacuum environments where access is limited. 

\endgroup

\newpage
\section*{ACKNOWLEDGEMENTS}
This work was supported by the U.K. Engineering and Physical Sciences Research Council via the EPSRC Hub in Quantum Computing and Simulation (EP/T001062/1), the U.K. Quantum Technology hub for Networked Quantum Information Technologies (No. EP/M013243/1), the European Commission’s Horizon-2020 Flagship on Quantum Technologies Project No. 820314 (MicroQC), the U.S. Army Research Office under Contract No. W911NF-14-20106 and Contract No. W911NF-21-1-0240, the Office of Naval Research under Agreement No.N62909-19-1-2116, and the University of Sussex.

\section*{AUTHOR DECLARATIONS}
\subsection*{Conflict of interest}
The authors have no conflicts to disclose.

\subsection*{Author Contributions}
\textbf{Toby Maddock}  Conceptualization (equal); Formal analysis (equal); Investigation (equal); Validation (lead); Visualization (equal); Writing – original draft (lead); Writing – review  editing (equal). \textbf{Parsa Rahimi} Conceptualization (equal); Investigation (equal); \textbf{Matthew Aylett} Investigation (supporting); Resources (lead). \textbf{Rares Barcan} Investigation (supporting); Writing – original draft (supporting); Writing – review  editing (supporting). \textbf{Sebastian Weidt} Supervision (equal). \textbf{Winfried Karl Hensinger} Supervision (equal); Writing - Review and editing (equal); Conceptualization (equal); Funding acquisition (equal).

\section*{DATA AVAILABILITY}
The data that support the findings of this study are available
from the corresponding author upon reasonable request.

\section{REFERENCES}
\bibliography{hehe}
\end{document}